\documentclass[reprint,pdflatex,
nofootinbib,
amsmath,amssymb,
aps,
floatfix,showpacs
]{revtex4-2}

\usepackage{graphicx}
\usepackage{dcolumn}
\usepackage{bm}
\usepackage{lineno}
\usepackage{float}
\usepackage{xcolor}
\usepackage{appendix}
\usepackage{hyperref}
\usepackage{mathrsfs}
\usepackage{booktabs}
\usepackage{array}
\usepackage{multirow}

\definecolor{mygreen}{rgb}{0., 0.5, 0.}

\newcommand{\dimensionlesspr}{\hat{P}}
\newcommand{\dimensionlessnu}{\hat{\mathcal{N}}}
\newcommand{\dimensionlessch}{\hat{\chi}}

\newcommand{\dimensionlesssusctwo}{\hat{\chi}_2^B}
\newcommand{\dimensionlesssuscfour}{\hat{\chi}_4^B}
\newcommand{\curtwo}{\kappa_2^B}

\newcommand{\curjusttwo}{\kappa_2}

\newcommand{\partitionfunction}{\mathcal{Z}}
\newcommand{\hmu}{\hat{\mu}}

\newcommand{\hmuB}{\hat{\mu}_B}

\newcommand{\muB}{\mu_B}

\newcommand{\LA}{\left \langle}
\newcommand{\RA}{\right \rangle}

\newcommand{\Ob}{\mathcal{O}}

\newcommand{\pade}{Pad\'e~}

\newcommand{\pseudocriticalresum}{\text{T}_{\text{pc}}^{\text{R}}}

\newcommand{\curtwoscl}{\kappa_{2,{\rm sc}}^B}
\newcommand{\curtwocrt}{\kappa_{2,{\rm pc}}^B}

\newcommand{\curjusttwoscl}{\kappa_{2,{\rm sc}}}
\newcommand{\curjusttwocrt}{\kappa_{2,{\rm pc}}}

\newcommand{\pseudocritical}{{\rm T}_{\rm pc}}

\usepackage{float}
\usepackage{calc}
\newlength{\depthofsumsign}
\begin{document}

\preprint{APS/123-QED}

\title{QCD crossover temperature and curvature coefficient from unbiased exponential resummation at physical quark masses in (2+1)-flavor lattice QCD}

\author{Sabarnya Mitra}
\email{smitra@physik.uni-bielefeld.de}
\affiliation{Fakult\"at f\"ur Physik, Universit\"at Bielefeld, D-33615 Bielefeld,
Germany}

\date{\today}

\begin{abstract}

We present the first application of the unbiased exponential
resummation (UER) method to the determination of the QCD pseudocritical
temperature $\pseudocritical$ at vanishing baryon chemical potential
$\muB$. By reconstructing the finite-density baryon number
susceptibility $\chi_2^B$, we define two thermal-derivative
observables whose peak positions yield
$\pseudocritical^T=160.2(4)(3)$ MeV and
$\pseudocritical^C=158.1(5)(3)$ MeV. We also show that the pseudocritical temperature, $\pseudocritical^\chi$
obtained from the peak of the fourth-order baryon number susceptibility
$\chi_4^B$ is consistent with these determinations and
with previous lattice-QCD results based on Taylor expansions
\cite{Bollweg:2022Pade}. Further, we demonstrate that scaling relations \cite{Bollweg:2022fqq} between $\chi_4^B$ and the thermal derivatives of
$\chi_2^B$ yield mutually consistent estimates of the
leading second order curvature coefficient $\curtwo$ at the corresponding
pseudocritical temperatures. A direct analysis of the $\muB$ dependence of the pseudocritical temperatures provides an independent but substantially less constrained determination of the curvature, which remains statistically consistent with the scaling-based estimates. These results demonstrate the consistency of this UER method with the expected scaling behavior and establish it as a complementary approach for probing the QCD crossover at small finite baryon density.

\end{abstract} 
 

\maketitle

\section{Introduction}
\label{sec:Introduction}

Understanding the thermodynamics of strongly interacting matter at
non-vanishing temperature and baryon chemical potential $\muB$ is one
of the central goals of finite-density Quantum Chromodynamics (QCD).
In particular, determining the location and curvature of the QCD
crossover line in the vicinity of $\muB=0$
\cite{Allton:2002zi,Allton:2005gk,HotQCD:2014kol,HotQCD:2018baz,Borsanyi:2020fev}
is important for constraining the QCD phase diagram and interpreting
heavy-ion collision experiments. Since lattice QCD calculations at
finite $\muB$ are hindered by the sign problem, most first-principles
Monte Carlo studies have relied on expansions about $\muB=0$, such as
Taylor expansions of thermodynamic observables
\cite{Gupta:2004pk}, or on analytic continuation from simulations at
imaginary $\muB$ \cite{deForcrand:2002hgr}, where the sign problem is
absent.

Recently, the framework of unbiased exponential resummation (UER)
\cite{Mitra:2022cum,Mitra:2023unb,Mitra:2023phd,Mitra:2022Lat,Mitra:2022dae}
has been developed as an approach for studying finite-density QCD
thermodynamics. Its first direct application to the study of Lee-Yang zeros of the QCD partition function at
finite isospin density has been demonstrated in Ref.~\cite{Mitra:2024czm}.
By exponentiating a Taylor-like expansion truncated at a given order
in $\muB$, the formalism constructs a finite-density representation of
the QCD partition function $\partitionfunction$, that reproduces the 
Taylor expansion exactly up to the truncation order in $\muB$ remaining in the exponential argument while
generating higher-order terms through the exponential structure.
This provides a way to reconstruct thermodynamic observables over a
finite range of $\muB$\footnote{In principle, this can be generalised to any chemical potential.} without explicitly computing all
higher-order Taylor coefficients. Previous studies have established
and benchmarked this framework for finite-density thermodynamics
\cite{Mitra:2022cum,Mitra:2023unb,Mitra:2023phd}, while its application
to pseudocritical observables has remained unexplored. In particular,
whether the finite-density information reconstructed through unbiased
exponential resummation can be used to consistently determine the pseudocritical line at small $\muB$ including the estimates of $\pseudocritical$ and $\curtwo$, has not yet been established.

In this paper, we address this question by applying UER to the determination of the QCD pseudocritical
line at small baryon chemical potential, with
$\mu_Q=\mu_S=0$\footnote{Here, $\mu_Q$ and $\mu_S$ denote the
electric-charge and strangeness chemical potentials, respectively.},
similar to Ref.~\cite{Bollweg:2022Pade}. Using $(2+1)$-flavor lattice
QCD simulations with physical quark masses on a $32^3\times8$ lattice,
we reconstruct the baryon number susceptibility
$\chi_2^B$ and determine the pseudocritical temperature
from the peak positions of its thermal derivatives. We also estimate the same from the peak of fourth order baryon susceptibility $\chi_4^B$.  We further
determine the leading-order curvature coefficient $\curtwo$ using
scaling relations involving $\chi_4^B$ and the thermal
derivatives of $\chi_2^B$, and independently probe the
curvature through the explicit $\muB$ dependence of the
pseudocritical temperatures. Throughout this paper, we denote
$\hmuB\equiv\muB/T$.

 
\section{UER and lattice setup}
\label{sec:formalism-and-setup}

\subsection{Formalism in brief}

\begin{figure*}[ht]
    \begin{minipage}[t]{0.99\textwidth}
    \centering   
    \includegraphics[width=.328\textwidth]{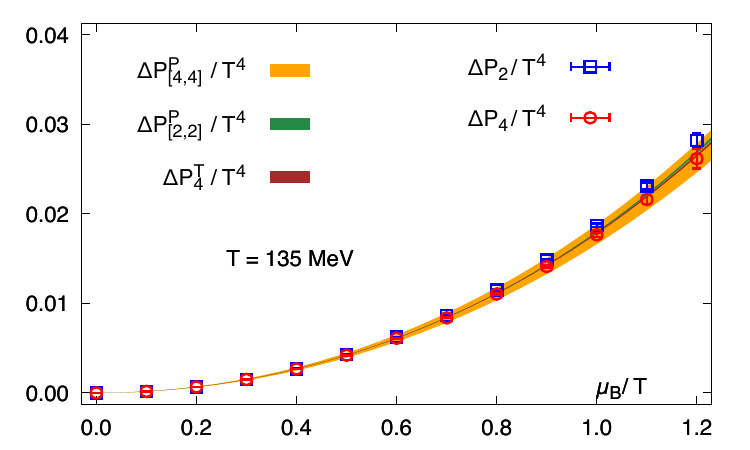}
    \includegraphics[width=.328\textwidth]{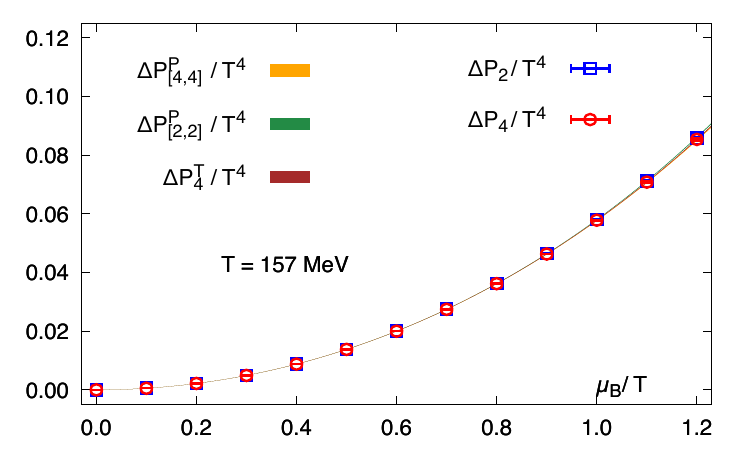}    
    \includegraphics[width=.328\textwidth]{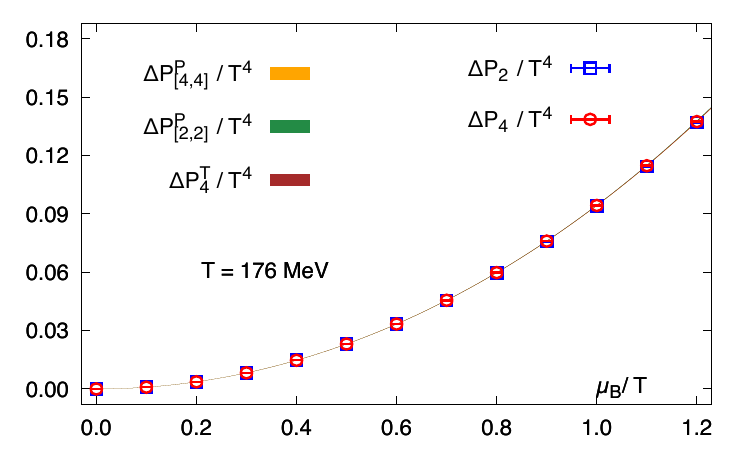}
    \includegraphics[width=.328\textwidth]{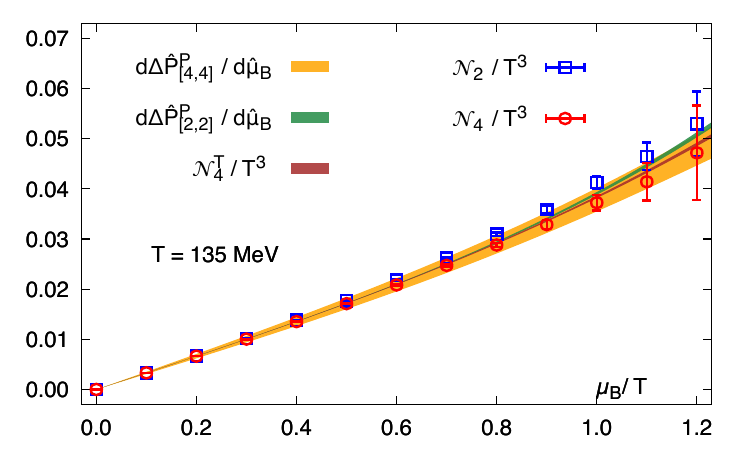}
    \includegraphics[width=.328\textwidth]{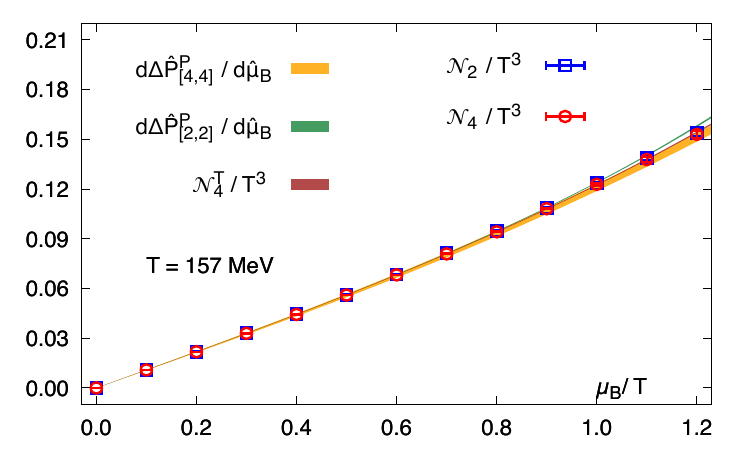}
    \includegraphics[width=.328\textwidth]{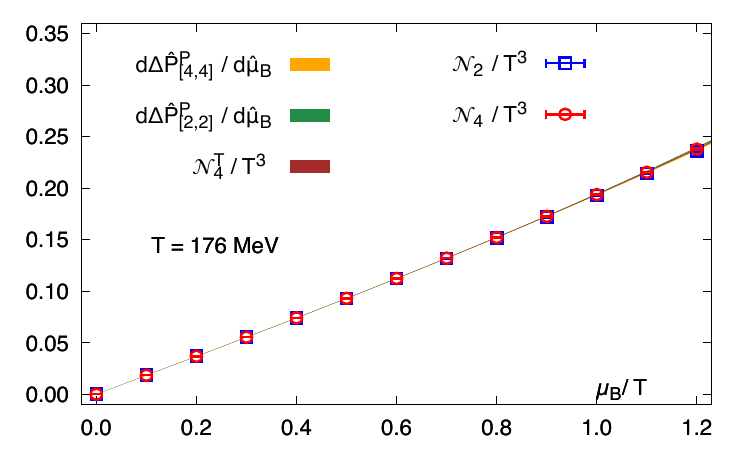}     
    \caption{Comparison of results between UER (blue and red points) and same obtained from Pad\'e resummation (orange and green bands) and Taylor expansion results (brown bands) for
    excess pressure $\Delta \hat{P}_{4}$ (top) and
    number density $\dimensionlessnu_{3}$ (bottom), where direct derivative results $\partial \dimensionlesspr_{[2,2]}^P / \partial \hmuB$ are also presented (magenta bands). 
    }
    \label{fig:exponential-vs-Pade}
    \end{minipage}
\end{figure*}

The method of UER \cite{Mitra:2022cum,Mitra:2023unb,Mitra:2023phd} provides a representation of the ratio, $R_N$ of $\partitionfunction$ at finite $\hmuB$ to $\partitionfunction$ at zero $\hmuB$ truncated to $N^{\rm th}$ order in $\hmuB$, by means of an average over a gauge ensemble simulated at $\hmuB=0$, 
\begin{equation}
    R_N=\frac{\partitionfunction_N(\hmuB)}{\partitionfunction(0)} = \LA\, \exp{\left[\sum_{n=1}^N\,\frac{\overline{D}_n}{n!}\,\hmuB^n\right]}\,\RA \;.
    \label{eq:unbiased-exp-resum}
\end{equation}
Here, $\overline{D}_n$ is the $n^{\rm th}$ order $\hmuB$ derivative of the logarithm of the fermion determinant at $\hmuB=0$. This includes the correction terms  \cite{Mitra:2023unb} required to ensure the unbiased nature of the exponential resummation procedure. At a finite temperature $T$, the excess pressure $ \dimensionlesspr=\Delta P/T^4$, number density $\dimensionlessnu=\mathcal{N}/T^3$, baryon number susceptibility $\dimensionlesssusctwo=\chi_2^B/T^2$ to $\Ob(\hmuB^N)$ are given by,
\begin{equation}
    \dimensionlesspr_N=\frac{\ln R_N}{VT^3} \;\;\;,\;\;\;\dimensionlessnu_N = \frac{\partial \dimensionlesspr_N}{\partial \hmuB}\;\;\;,\;\;\; \dimensionlessch_{2,N}^B = \frac{\partial^2 \dimensionlesspr_N}{\partial \hmuB^2},
    \label{eq:cumulants}
\end{equation}
where $V$ is the system volume.    
In this paper, we apply this formalism using $N=4$ and evaluate  $\dimensionlessch_{2,4}^B$ in Eq.\ref{eq:cumulants} derived from the  stochastic estimate of $R_4$ obtained here using bootstrap sampling. This reproduces the Taylor expansion up to $\Ob(\hmuB^2)$ with correction terms appearing from $\Ob(\hmuB^4)$ onwards.
Subsequent statistical uncertainties are obtained taking into account the correlation terms appearing in the covariance matrix. 
With the data points of $\dimensionlessch_{2,4}^B$ as function of $T$, we proceed on determining the estimates of $\pseudocritical$ and $\curtwo$ in this paper. A comparative study of these estimates obtained from $N=2$ and $N=4$ of UER is provided in Appendix \ref{appsec:systematic-err-an}. For simplicity, we denote  $\dimensionlessch_{2,4}^B$ as $\dimensionlesssusctwo$ from here on.

\subsection{Lattice setup}

We perform $(2+1)$-flavor lattice QCD simulations using the Highly Improved Staggered Quark (HISQ)
action with physical light and strange-quark masses on
$32^3\times 8$ lattices over the temperature range
$T\in[135:176]$ MeV, with $\mathcal{O}(10^5\!-\!10^6)$ gauge
configurations at each temperature. Along with the HISQ action, we consider the $\Ob(a^2)$ improved Symanzik gauge action \cite{Symanzik:1983dc} together which constitute the QCD action, $S_{\rm QCD}$. \footnote{This QCD action $S_{\rm QCD}$ enters as Boltzmann weight in QCD partition function integral, $Z=\int DU \,\exp{(-S_{\rm QCD})}$.} The scale setting of this work determining the different input $T$ values of the $\dimensionlessch_2$ data points, is based on the Line of Constant Physics (LCP) given in Ref.~\cite{HotQCD:2014kol}.
The ratio $R_4$ is evaluated
using the full configuration ensemble at each temperature, with
bootstrap sampling employed to estimate the statistical
uncertainties. Using this setup and the same bootstrap sampling framework, we likewise estimate $\pseudocritical$ and $\curtwo$, along with associated uncertainties of these observables.

\section{Comparison with \pade results}
\label{sec:exponential-vs-Pade}

Prior to discussing our main results of pseudocritical temperature and curvature coefficient estimates evaluated using UER, it is important to review and re-check the reliability of these resummation results of bulk thermodynamic observables. Although done before in the context of Taylor series expansion in Ref.\cite{Mitra:2023unb}, we show this here by comparing these results with similar results derived from \pade resummation \cite{Bollweg:2022Pade}. 

We present this comparison in Fig.\ref{fig:exponential-vs-Pade}, where the second and fourth order unbiased resummation results for excess pressure $\dimensionlesspr$, and number density $\dimensionlessnu$ are outlined alongside similar and higher order \pade results as function of $\hmuB$, for $T=135,157$ and $176$ MeV. While we use \pade [2,2] and [4,4] results for $\dimensionlesspr$, their spline $\hmuB$-derivatives are considered as similar \pade estimates for $\dimensionlessnu$.
Almost two orders of higher statistics as compared to Ref.\cite{Mitra:2023unb} have been used for these new UER results. 
Besides having agreement between $N=2$ and $N=4$ UER results of $\dimensionlesspr$ and $\dimensionlessnu$ in this $\hmuB$ range, equally apparent from this figure, is the promising consistency between these two resummation procedures. These results are also well-consistent with similar order Taylor series values shown here in Fig.\ref{fig:exponential-vs-Pade}. In this section, we have restricted this discussion and associated comparative analysis for $\hmuB \in $ [0\;:\;1.2]  only. This is adequate enough to support the reliability of UER results needed and relevant for this work, where the central estimates of $\pseudocritical$ and $\kappa_2^B$ are evaluated by making use of these UER results in small $\hmuB$ regime, $\hmuB \leq 0.5$ only. 
A further study in higher baryon density along this direction for more temperatures, is reserved for a future work.

\begin{figure}
    \centering
    \includegraphics[width=.45\textwidth]{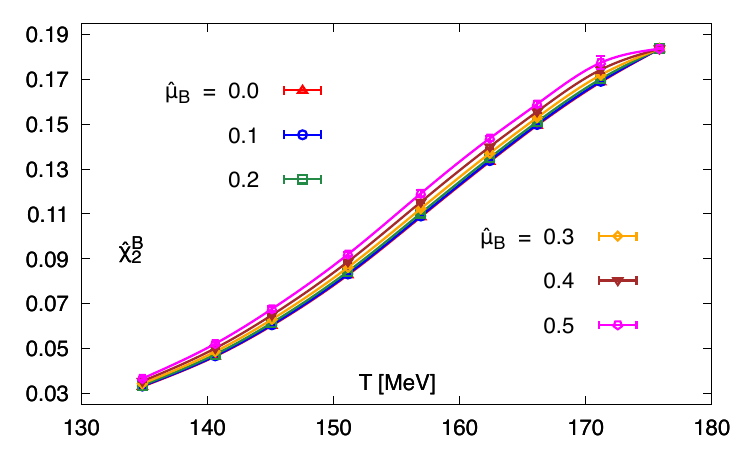}  
    \caption{The baryon number susceptibility, $\dimensionlessch_2^B$ obtained using UER on $N_\tau=8$ lattices, shown as a function of $T$, for $0 \leq \hmuB \leq 0.5$. Six different colors are used for the six values of $\hmuB$. (More details in text)}
    \label{fig:chi2vsmuB}
\end{figure}

\section{Pseudocritical temperature}
\label{sec:pseudocritical}

\begin{figure}
     \centering
        \includegraphics[width=.41\textwidth]{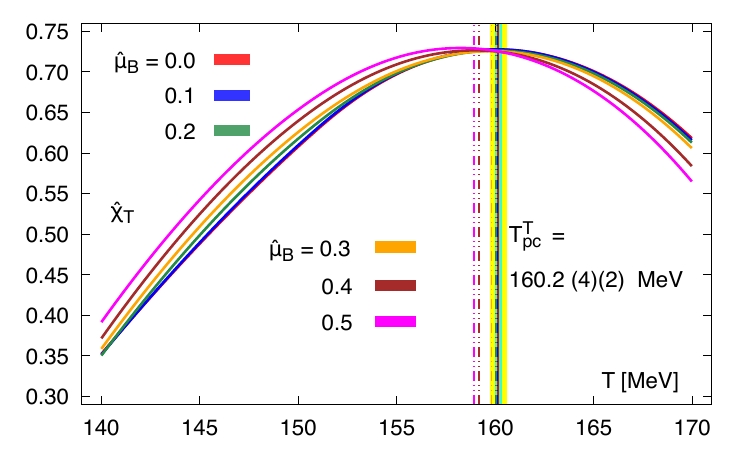}  \\
         \includegraphics[width=.41\textwidth]{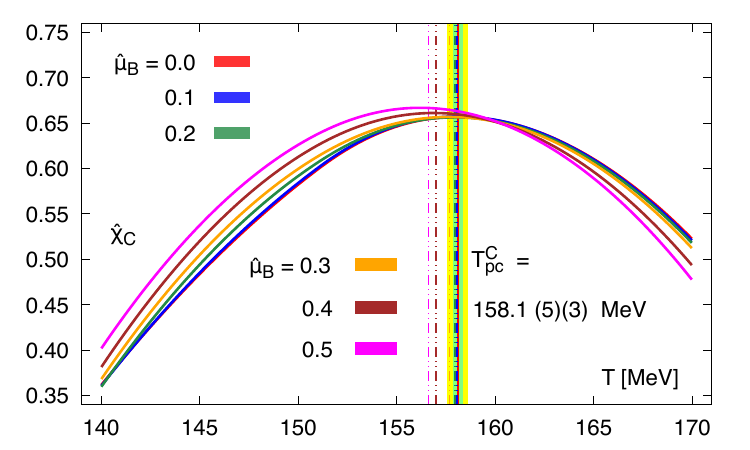} 
     \caption{\hspace{-.41cm} 
     ({\it Top:}) The thermal derivative, $\dimensionlessch_T$ (Eq.\ref{eq:thermal derivatives}) vs $T$ shown for the same six $\hmuB$ values as in Fig.\,\ref{fig:chi2vsmuB}, following the same color convention.
     ({\it Bottom:}) The corresponding result for $\dimensionlessch_C$ (Eq.\ref{eq:thermal derivatives}). The peak positions for non-vanishing $\hmuB$ are indicated by vertical dotted lines, with similar color convention. The red solid line gives the same for $\hmuB=0$, thereby giving  estimates of  $\pseudocritical^T,\pseudocritical^C$. The yellow (cyan) bands show the statistical (systematic) errors on respective  $\pseudocritical$ values.
     }
     \label{fig:chi2B4thord-vs-T}
\end{figure} 

An important purpose of this work, as discussed before is determination of pseudocritical temperature $\pseudocritical$ from UER. This is done by studying the temperature dependence of the baryon
number susceptibility, $\dimensionlesssusctwo$, and subsequently considering
its temperature derivative. As already argued in Ref.~\cite{Bollweg:2022fqq},
the temperature derivative of $\dimensionlesssusctwo$ develops a finite
pronounced peak at the chiral transition temperature $T_c$ in the chiral
limit, which broadens away from this limit.
At physical quark masses, this broadened peak therefore
provides an estimator of the pseudocritical temperature $\pseudocritical$.
We therefore estimate $\pseudocritical$ from the peak positions of two such
corresponding derivative observables at physical light-quark masses, at vanishing $\hmuB$, as discussed in the following sections.

\subsection{From \texorpdfstring{$\chi_2$}{chi2} analysis}
The temperature dependence of the baryon number susceptibility,
$\dimensionlesssusctwo$, obtained using the unbiased exponential
resummation framework on $N_\tau=8$ lattices, is shown in
Fig.~\ref{fig:chi2vsmuB} for six values of $\hmuB$,
$\hmuB\in[0\,:\,0.5]$. We construct smooth cubic splines to interpolate
the resummed data points of $\dimensionlesssusctwo$ as a function of
temperature. We then subsequently construct two thermal derivative observables,
\begin{equation}
\dimensionlessch_T
=
T\,\frac{\partial\dimensionlessch_2^B}{\partial T}
\;\;\;,\quad
\dimensionlessch_C
=
T_c\,\frac{\partial\dimensionlessch_2^B}{\partial T},
\label{eq:thermal derivatives}
\end{equation}
whose maxima define the corresponding estimates
$\pseudocritical^T$ and $\pseudocritical^C$ of the pseudocritical
temperature $\pseudocritical$ at $\hmuB=0$.

For the definition of $\dimensionlessch_C$ in Eq.\ref{eq:thermal derivatives}, we take
$T_c=145$ MeV \cite{Mitra:2024mke,Mitra:2025aeu,Mitra:2025hsk,Mitra:2026bbg}.
As discussed above, $\pseudocritical$ approaches $T_c$ in the vanishing
light-quark-mass limit, consistent with the expected universal scaling
behavior of the chiral transition \cite{Ding:2024sux}.
The corresponding observables $\dimensionlessch_T$ and
$\dimensionlessch_C$, shown in Fig.~\ref{fig:chi2B4thord-vs-T}, are
obtained analytically as functions of $T$ from the spline interpolation.
The interpolation range is restricted to $[135:172]$ MeV, which
encompasses the crossover region while avoiding poorly constrained
extrapolation of the constructed cubic splines.

Fig.~\ref{fig:chi2B4thord-vs-T} illustrates that both
$\dimensionlessch_T$ and $\dimensionlessch_C$ exhibit clear peak
structures as functions of $T$. Consistent with Ref.~\cite{Bollweg:2022fqq},
we find that the peak positions shift toward lower temperatures, while
the corresponding peak heights increase with increasing $\hmuB$ although, the latter increase is incremental over this range of $\hmuB$ considered. At
$\hmuB=0$, the peak positions of these two observables yield
\begin{equation}
\pseudocritical^{T}=160.2(4)(2)\;\,{\rm MeV},\quad
\pseudocritical^{C}=158.1(5)(3)\;\,{\rm MeV},
\label{eq:Tpc from chi_T and chi_C}
\end{equation}
where the first and second uncertainties denote statistical and
systematic errors, respectively. The difference between
$\pseudocritical^T$ and $\pseudocritical^C$ is consistent with the
expected observable dependence of the crossover temperature at physical
quark masses.
Although the same thermal derivatives exhibit well-defined peaks also
at finite $\hmuB$, their interpretation as estimators of the
finite-density pseudocritical line requires a dedicated analysis of the
underlying universal scaling behavior. We therefore restrict the present
determination of $\pseudocritical$ to $\hmuB=0$ and leave a detailed
finite-density analysis of possible pseudocritical temperature estimates
for future work. 

\subsection{From \texorpdfstring{$\chi_4$}{chi4} analysis}

We further explore the temperature dependence of the fourth-order
baryon number susceptibility, $\dimensionlesssuscfour$, as a complementary estimator of $\pseudocritical$ at $\hmuB=0$.
As the second derivative of $\dimensionlesssusctwo$ with respect to
$\hmuB$, $\dimensionlesssuscfour=\partial^2 \dimensionlesssusctwo / \partial \hmuB^2$ at $\hmuB=0$ exhibits a pronounced peak in the
crossover region, providing a similar determination of the
pseudocritical temperature $\pseudocritical$.
Like the thermal derivative observables
$\dimensionlessch_T$ and $\dimensionlessch_C$, the interpretation of
the peak of $\dimensionlesssuscfour$ at finite $\hmuB$ involves
additional higher-order derivatives of the universal scaling
functions and is therefore beyond the scope of the present analysis.

\begin{figure}[t]
    \centering
    \includegraphics[width=.47\textwidth]{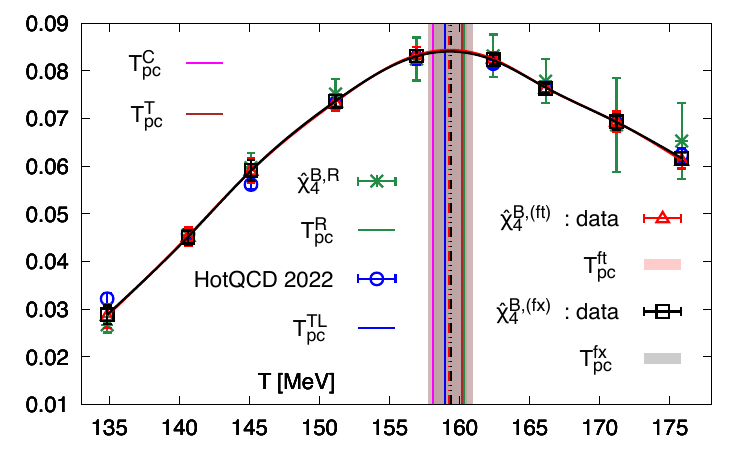}
    \caption{$\dimensionlessch_4^B$ calculated from Eq.\ref{eq:chi4B-vs-muB} shown as a function of $T$ using {\it i)} $\dimensionlessch_2^B(0)$ directly from the resummed data (red triangles) and {\it ii)} treating $\dimensionlessch_2^B(0)$ as a fit parameter (black squares). The red and black bands outline $\pseudocritical^{\rm ft}$ and $\pseudocritical^{\rm fx}$ with errors. The magenta and brown solid lines highlight $\pseudocritical^C$ and $\pseudocritical^T$. The green and blue points (lines) outline $\dimensionlesssuscfour$ ($\pseudocriticalresum$) obtained from directly resummed result and HotQCD Taylor analyses ($\pseudocritical^{\rm TL}$) \cite{Bollweg:2022Pade}, respectively.
    }
    \label{fig:chi4vsT}
\end{figure}

We obtain $\dimensionlesssuscfour(T)$ at $\hmuB=0$ from the
$\hmuB$ dependence of the resummed $\dimensionlesssusctwo$ at each
temperature,
\begin{equation}
    \dimensionlessch_2^B(\hmuB)
    =
    \dimensionlessch_2^B(0)
    +\frac{\hmuB^2}{2!}\,\dimensionlessch_4^B(0)
    +\Ob(\hmuB^4)\;.
    \label{eq:chi4B-vs-muB}
\end{equation}
The resulting $\dimensionlesssuscfour(T)$ is shown in
Fig.~\ref{fig:chi4vsT}, where it exhibits good agreement with the
corresponding Taylor-expansion results produced by HotQCD collaboration\, \cite{Bollweg:2022Pade},
shown as blue points for each temperature in this figure.
The peak positions of our $\dimensionlessch_4^{B,{\rm (ft)}}$ and $\dimensionlessch_4^{B,{\rm (fx)}}$ estimates yield
\begin{equation}
 \pseudocritical^{\rm ft}=159.4(10)(12)\ {\rm MeV},
 \quad
 \pseudocritical^{\rm fx}=159.4(9)(8)\ {\rm MeV}\,.
 \label{eq:Tpc-from-chi4}
\end{equation}
In the ``ft" determination, $\dimensionlesssusctwo(0)$ is treated as a fit parameter simultaneously together with $\dimensionlesssuscfour(0)$ whereas in the ``fx" determination, $\dimensionlesssusctwo(0)$ is fixed to its directly reconstructed resummed value, for each temperature.

We show these results in Fig.\,\ref{fig:chi4vsT}, in which we find that both determinations in Eq.~\ref{eq:Tpc-from-chi4} are consistent
with each other and exhibit good agreement with the pseudocritical
temperature estimates $\pseudocritical^T$ and $\pseudocritical^C$
obtained from the thermal derivative observables
$\dimensionlessch_T$ and $\dimensionlessch_C$ in
Eq.~\ref{eq:Tpc from chi_T and chi_C}, as shown in
Fig.~\ref{fig:chi2B4thord-vs-T}.
We additionally evaluate $\dimensionlesssuscfour$ directly within
the UER framework, and show the first result, $\dimensionlessch_4^{B,{\rm R}}$ from this setup in Fig.\,\ref{fig:chi4vsT}. This observable, at $\hmuB=0$, shows consistent behavior with other results and yields
\begin{equation}
    \pseudocriticalresum=160.3(40)(13)\ {\rm MeV}.
\end{equation}
The larger statistical uncertainty reflects the enhanced noise and
correlations associated with the direct computation of a  
higher-order observable, within the bootstrap analysis of the UER framework. A detailed study of its statistical and
systematic uncertainties is left for future work.
For comparison, the maximum of $\dimensionlesssuscfour$ obtained from
the Taylor-expansion analysis~\cite{Bollweg:2022Pade} using
the same interpolating procedure, gives a consistent estimate,

\begin{equation}
\pseudocritical^{\rm TL}=158.9(10)(14)\ {\rm MeV}.    
\end{equation}
%

\section{Curvature coefficient}
\label{sec:curvature}

\subsection{Scaling estimate}

At $\hmuB=0$, the leading-order curvature coefficient $\curtwo$
can be extracted from ratios of the singular contributions to the
relevant observables, as demonstrated in Ref.~\cite{Bollweg:2022fqq}.
Since our analysis is performed within the scaling regime, we neglect
the sub-leading non-singular contributions. In the resulting ratios,
the non-universal scaling parameters and the corresponding derivatives
of the free-energy scaling function $f_f(z)$ cancel, allowing the
curvature coefficient to be expressed directly in terms of the
susceptibilities. We consider the following estimator for curvature,
\begin{equation}
\curtwoscl
= \frac{1}{6}\,
\frac{\dimensionlessch_4^B}{\dimensionlessch_C}\;,
\label{eq:kappa2-scaling}
\end{equation}
derived from the scaling relations, where
$\dimensionlessch_C=T_c\,d\dimensionlesssusctwo/dT$ is the
corresponding thermal derivative introduced above. One can also obtain this value from $\dimensionlessch_T$, using the identity $\dimensionlessch_T = T\,\dimensionlessch_C\,/\,T_c$.

We evaluate this curvature estimator using the two determinations
of $\dimensionlesssuscfour$ discussed above: (i) the result obtained
by treating $\dimensionlesssusctwo$ as a fit parameter, denoted by
$(\mathrm{ft})$, and (ii) the result obtained with
$\dimensionlesssusctwo$ fixed directly from the resummed data, denoted
by $(\mathrm{fx})$. The resulting values of
$\curjusttwoscl^{B,T}=\curtwoscl(\pseudocritical^T)$ and $\curjusttwoscl^{B,C}=\curtwoscl(\pseudocritical^C)$, evaluated at the
pseudocritical temperature estimates $\pseudocritical^T$ and
$\pseudocritical^C$ determined from the corresponding thermal
derivatives, are summarized in Table~\ref{tab:comparison}.

\begin{table}[H]
    \centering
    \setlength{\tabcolsep}{10pt}
    \renewcommand{\arraystretch}{1.9}
    \begin{tabular}{|c||c|c|}
    \hline
    Curvature & $\curtwoscl(\pseudocritical^T)$ & $\curtwoscl(\pseudocritical^C)$ \\
    \hline
      $\dimensionlessch_4^{B,{\rm (ft)}}$  & $0.02082\,(18)$ & $0.02062\,(22)$  \\
      \hline
      $\dimensionlessch_4^{B,{\rm (fx)}}$  & $0.02083\,(17)$ & $0.02066\,(21)$ \\
      \hline
    \end{tabular}
    \caption{Comparison of the second order curvature estimate, $\curtwo$, at the two pseudocritical temperatures, $\pseudocritical^T=160.2$ MeV and $\pseudocritical^C=158.1$ MeV, as mentioned in Eq.\ref{eq:Tpc from chi_T and chi_C}.}
    \label{tab:comparison}
\end{table}

For $T=\pseudocritical^T=160.2$ MeV, we obtain
$0.02082(18)$ and $0.02083(17)$ using the
$\dimensionlessch_4^{B,{\rm (ft)}}$ and
$\dimensionlessch_4^{B,{\rm (fx)}}$ determinations, respectively.
The corresponding $\curjusttwoscl^{B,T}$ and $\curjusttwoscl^{B,C}$
estimates agree within the quoted precision. At
$T=\pseudocritical^C=158.1$ MeV, the corresponding values are
$0.02062(22)$ and $0.02066(21)$. Thus, the agreement between the values of the $\curtwo$ estimates in $(\mathrm{ft})$ and $(\mathrm{fx})$ determinations provides an additional consistency check on the scaling analysis.

Averaging the corresponding estimates at each pseudocritical
temperature using bootstrap analysis on each of the working 2000 bootstrap samples gives
\begin{equation}
\curjusttwoscl^{B,T} = 0.02082 \,(14) \quad ,\quad 
\curjusttwoscl^{B,C} = 0.02064 \,(15)\,.
\label{eq:kappa2values-from-scalingrelations}
\end{equation}
These values are consistent with the curvature obtained at
comparable temperatures from Taylor expansion methods
\cite{Bollweg:2022fqq}. Their small difference is also consistent
with the quoted uncertainties, indicating that the scaling-based
curvature estimate is stable over the range of pseudocritical
temperatures considered here.

\begin{table}[htbp]
    \centering
    \setlength{\tabcolsep}{10pt}
    \renewcommand{\arraystretch}{1.9}
    \begin{tabular}{|c||c|}
    \hline
    Curvature & $\curtwoscl(\pseudocritical^\chi)$ \\
    \hline
      $\dimensionlessch_4^{B,{\rm (ft)}}$  & $0.02073\,(20)$ \\
      \hline
      $\dimensionlessch_4^{B,{\rm (fx)}}$  & $0.02076\,(18)$ \\
      \hline
    \end{tabular}
    \caption{Curvature estimate $\curtwo$ at $\pseudocritical^\chi=159.4$ MeV, as obtained from the peak position of $\dimensionlesssuscfour$ in $T$ (Eq.\ref{eq:Tpc-from-chi4}).}
    \label{tab:comparisonatTpcchi4}
\end{table}

We further perform the same analysis at
$T=\pseudocritical^\chi=159.4$ MeV where,
$\pseudocritical^\chi$\footnote{Here, $\pseudocritical^\chi$ is obtained as the bootstrap average of (ft) and (fx) determinations introduced before.} is determined from the peak position of
$\dimensionlesssuscfour$ as a function of temperature. The
corresponding curvature estimates are shown in
Table~\ref{tab:comparisonatTpcchi4}. The values obtained from
$\curjusttwo^{B,T}$ and $\curjusttwo^{B,C}$, as well as those from
the $(\mathrm{ft})$ and $(\mathrm{fx})$ determinations, are again
consistent. Averaging these estimates gives
\begin{equation}
\curtwoscl(\pseudocritical^\chi) = 0.02075\,(14)\,.
\label{eq:kappa2values-from-chi4}
\end{equation}
All central values and uncertainties quoted in
Eqs.~\ref{eq:kappa2values-from-scalingrelations} and
\ref{eq:kappa2values-from-chi4} are obtained from bootstrap
analysis.
The correlations among these derived observables are taken into account through the bootstrap analysis. All quantities are evaluated on the same bootstrap samples, such that their statistical correlations are retained when forming temperature derivatives, ratios, and pseudocritical temperature as well as curvature coefficient estimates.
In particular, $\dimensionlessch_T$ and $\dimensionlessch_C$ are obtained from the temperature dependence of $\dimensionlessch_2^B$ within each bootstrap sample, while the curvature estimates involving $\dimensionlessch_4^B/\dimensionlessch_T$ and $\dimensionlessch_4^B/\dimensionlessch_C$ are likewise evaluated on each bootstrap sample before determining the statistical distribution of the resulting quantity. Thus, the quoted uncertainties include and do take into account, the correlations between the underlying observables rather than treating their statistical fluctuations as independent.

When taken together along with this error analysis, we observe the three estimates
$\curtwoscl(\pseudocritical^T)$,
$\curtwoscl(\pseudocritical^C)$, and
$\curtwoscl(\pseudocritical^\chi)$, as presented in Eqs.\ref{eq:kappa2values-from-scalingrelations} and \ref{eq:kappa2values-from-chi4} are mutually consistent. This
demonstrates the stability of $\curtwoscl$, the scaling-based curvature estimate of $\curtwo$ with respect to the choice of pseudocritical temperature estimator.

\subsection{From pseudocritical line analysis}

Next, we estimate $\curtwo$ from the $\hmuB$ dependence of the pseudocritical temperatures $\pseudocritical^T$ and $\pseudocritical^C$, determined from the peak positions of $\dimensionlessch_T$ and $\dimensionlessch_C$, respectively, as discussed in the previous section. 

At vanishing baryon chemical potential, \(\hmuB=0\), we determine the pseudocritical temperatures \(T_{pc}^{T}\) and \(T_{pc}^{C}\) from the peak positions of the thermal derivatives \(\dimensionlessch_T=T\,d\dimensionlessch_2^B/dT\) and \(\dimensionlessch_C=T_c\,d\dimensionlessch_2^B/dT\) respectively. This construction is motivated by the scaling behavior of \(\dimensionlessch_2^B\), which is an energy-like observable. While these observables $\dimensionlessch_T$ and $\dimensionlessch_C$ develop a cusp at chiral phase transition temperature $T_c$ in the chiral limit, this cusp gets smoothed into a finite broad peak for physical quark masses at pseudocritical temperatures $\pseudocritical^T$ and $\pseudocritical^C$. In the chiral limit, this cusp corresponds to the non-analytic singular behavior of chiral phase transition, which is a genuine phase transition \cite{Bollweg:2022Pade}. This non-analytic singular form of temperature dependence in the chiral limit gets smoothed into an analytic, finite maximum in its thermal derivative in the crossover region, away from the chiral limit.  Since different observables generally attain their maxima at different temperatures in a crossover, the peak positions of these derivative-based estimators define observable-dependent $\pseudocritical$ values, at $\hmuB=0$.

At finite nonzero \(\muB\), however this correspondence between the peak position of these estimators and the pseudocritical temperature estimates defined from the underlying universal scaling condition is not exact. In scaling terms, the leading singular temperature dependence of \(\dimensionlessch_2^B\) at \(\hmuB=0\) is governed entirely by the first derivative of the free-energy scaling function, \(f_f'(z)\). At finite \(\hmuB\) in contrast, an explicit chemical-potential dependence of the scaling variable enters in this relation, which generates additional contributions involving higher derivatives of the scaling function. Schematically upto $\Ob(\hmuB^2)$, the singular part of $\dimensionlessch_2^B$, $\dimensionlessch_2^{B,s}$ can be written as
\begin{equation}
\dimensionlessch_2^{B,s} (\hmuB)
\propto
\Big[f_f'(z(T,\muB))
+
K\,\hmuB^2\,f_f''(z(T,\muB))\Big],
\label{eq:singular-part-of-susc}
\end{equation}
where \(K\) is a positive non-vanishing proportionality factor containing the corresponding non-universal scaling factors.
Consequently taking a thermal derivative of $\dimensionlessch_2^{B,s}(\hmuB)$ generates contributions involving \(f_f''\) as well as leading higher order terms proportional to \(\hmuB^2 f_f'''\), which vanishes at $\hmuB=0$. Thus at finite \(\hmuB\), the peak of the thermal derivative of \(\chi_2^B\) does not in general, coincide exactly with the pseudocritical temperature defined by the underlying scaling condition. A naive quantitative extent of this approximation is briefed in Appendix \ref{appsec:pseudocritical-scaling-analysis}.

We therefore refer to the peak positions of \(\chi_T\) and \(\chi_C\) at finite \(\muB\) as susceptibility-based pseudocritical proxies, rather than identifying them exactly with the pseudocritical line. The distinction can become increasingly relevant with increasing \(\hmuB\), as higher-order finite-density contributions to the scaling observables become more important.
 However as detailed in Appendix \ref{appsec:pseudocritical-scaling-analysis}, this deviation is small within the restricted range $\hmuB\in$ [$0$\;:\;$0.5$] considered here for curvature extraction. In this context therefore, this allows us to use them as operational estimators of $\pseudocritical^T$ and $\pseudocritical^C$ in this section, and extract the corresponding curvature coefficients from respective fit analysis.
\begin{figure}[ht]
    \centering
    \includegraphics[width=.41\textwidth]{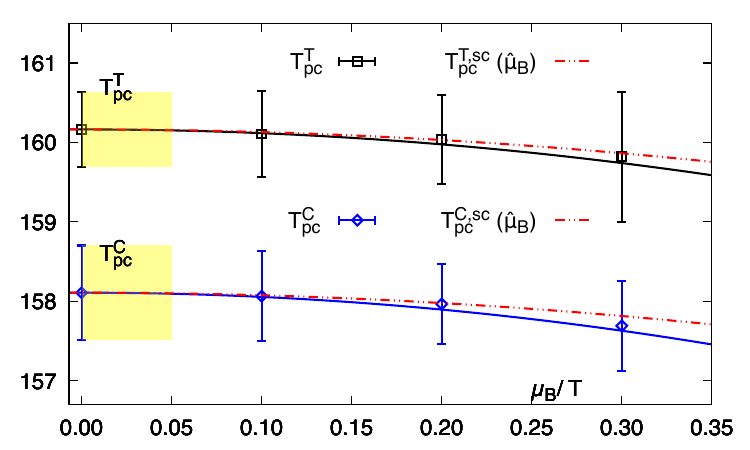}
    \caption{Estimates of $\curjusttwocrt^{B,T}$ and $\curjusttwocrt^{B,C}$ (see text for details) obtained from pseudocritical line analysis, using the parametrisation of $\pseudocritical(\hmuB)$ in Eq.\ref{eq:parametrisation of Tpc}. The yellow bands represent respective errors on $\pseudocritical^T$ and $\pseudocritical^C$ at $\hmuB=0$. (Details in text)
    }
    \label{fig:curvature-k2}
\end{figure}
For $x=T,C$, we parameterize the pseudocritical line as
\begin{equation}
\pseudocritical^x(\hmuB) =
\pseudocritical^x(0)\,\left[ 1-\kappa_{2,{\rm pc}}^{B,x}\,\hmuB^2 +\mathcal{O}(\hmuB^4)\right] ,
\label{eq:parametrisation of Tpc}
\end{equation}
where, we denote $\curjusttwocrt^{B,x} \equiv \curtwocrt(\pseudocritical^x)$ as the curvature estimate obtained from pseudocritical line in this section. We fit the $\hmuB$-dependent peak positions to this form over the range $0\leq\hmuB\leq 0.5$, obtaining
\begin{equation}
\curjusttwocrt^{B,T} = 0.029 (32)
\quad,\quad
\curjusttwocrt^{B,C} = 0.033 (27)\,.
\label{eq:curvature-from-Tpc}
\end{equation}
%
using only $\curjusttwocrt^{B,T}$ and $\curjusttwocrt^{B,C}$ as fit parameters in our fitting procedure. The solid black and blue curves outline the direct fit results of $\pseudocritical^T(\hmuB)$ and $\pseudocritical^C(\hmuB)$ of Eq.\ref{eq:parametrisation of Tpc}, while the dotted curves use instead the $\curtwo$ estimates of scaling relations, Eq.\ref{eq:kappa2values-from-scalingrelations}.
This comparison is illustrated in Fig.~\ref{fig:curvature-k2}, where we show only upto $\hmuB=0.3$ for visual representation. The
solid black and blue curves show the direct fits to the
$\pseudocritical^T(\hmuB)$ and $\pseudocritical^C(\hmuB)$ data using
Eq.~\ref{eq:parametrisation of Tpc}, with only the 
$\curjusttwo^{B,T}$ and $\curjusttwo^{B,C}$ as the fit parameters.
The dotted curves, $\pseudocritical^{T,{\rm sc}}(\hmuB)$ and $\pseudocritical^{C,{\rm sc}}(\hmuB)$ instead use the corresponding
$\pseudocritical^T(0)$ and $\pseudocritical^C(0)$ values from
Eq.~\ref{eq:Tpc from chi_T and chi_C}, together with the curvature
estimates obtained from the $\hmuB=0$ scaling relations in
Eq.~\ref{eq:kappa2values-from-scalingrelations}.

As shown in this figure, the two sets of curves are statistically consistent within the uncertainties of the pseudocritical line analysis. The central values of the curvature coefficients obtained from the direct analysis of the pseudocritical line in Eq.~\ref{eq:curvature-from-Tpc} are also found to be relatively close to the corresponding estimates from the scaling relations in Eq.~\ref{eq:kappa2values-from-scalingrelations}. The direct determination, however, remains less precisely constrained, as reflected by the comparatively large uncertainties on both $\curjusttwocrt^{B,T}$ and $\curjusttwocrt^{B,C}$. Thus, within the present precision, the two approaches yield mutually consistent estimates of the curvature.

\begin{table}[htbp]
    \centering
    \begin{tabular}{|c|| c | c | }
    \hline
    $\hmu_B^{max}$ & $\kappa_{2, {\rm pc}}^T$ & $\kappa_{2, {\rm pc}}^C$ \\
    \hline
       0.1 & 0.036 (317) & 0.028 (320)  \\
       \hline
       0.2 & 0.021 (80) & 0.023 (76) \\
       \hline
       0.3 & 0.023 (46) & 0.028 (35) \\
       \hline
       0.4 & 0.029 (36) & 0.033 (29) \\
       \hline
       0.5 & 0.029 (32) & 0.033 (27) \\
       \hline       
    \end{tabular}
\caption{Dependence of the curvature coefficients extracted from the pseudocritical line on the maximum value of $\hat\mu_B$ included in the fit. The pseudocritical temperatures are obtained using the cubic-spline interpolation over $135\leq T\leq172$ MeV, over the full range $0\leq\hat\mu_B\leq0.5$.}
    \label{tab:curvature-range}
\end{table}

The sensitivity of this direct determination of $\kappa_2^B$ to the fitting range, $\hmuB \in [0:\hmuB^{max}]$ is further illustrated in Table~\ref{tab:curvature-range}, where we show the extracted curvature coefficients, $\curjusttwocrt^T$ and $\curjusttwocrt^C$ for different values of the maximum chemical potential, $\hmuB^{max}$. In either case, we find the uncertainties decrease substantially as the fitting range is extended from $\hmuB^{max}=0.2$ to $0.3$, while the central values remain mutually consistent within the corresponding uncertainties. For $\hmuB^{max} > 0.3$, the extracted curvature coefficients show comparatively mild variations, with a further reduction of the uncertainties upon extending the range to $\hmuB^{max}=0.5$. This behavior reflects the finite-$\hmuB$ nature of the pseudocritical line analysis in which, a sufficiently large $\hat\mu_B$ lever arm is required to constrain the quadratic dependence, while the statistical fluctuations of the individual $T_{pc}(\hat\mu_B)$ determinations propagate into the global fit. The choice, $\hmuB^{max}=0.5$ therefore provides a substantially more precise determination while remaining consistent with the results obtained from the smaller fitting ranges.

In contrast, the scaling determination does not require a finite-$\mu_B$ fit and instead provide a local determination of the curvature from the observables and their temperature derivatives evaluated at $\mu_B=0$, resulting in smaller statistical uncertainties. Thus, although both approaches use the same underlying lattice data and determine the same physical curvature, they extract $\kappa_2^B$ through statistically different estimators. 
This comparison provides therefore a correlated internal cross-check of the curvature
coefficient $\curtwo$ using two complementary analyses based on shared ensembles. The scaling-relation
analysis determines $\curtwo$ entirely from $\hmuB=0$ susceptibilities,
following Ref.~\cite{Bollweg:2022fqq}, whereas the pseudocritical line
analysis probes the explicit $\hmuB$ dependence of the pseudocritical
temperatures $\pseudocritical(\hmuB)$. The statistical compatibility of the two determinations,
despite the substantially larger uncertainty and less-constrained nature of the direct determination, supports
the consistency of the curvature extracted from the scaling analysis.
This comparison is also consistent with the use of the
finite-volume-corrected value $T_c=145$ MeV
\cite{Mitra:2024mke,Mitra:2025aeu,Mitra:2025hsk,Mitra:2026bbg}
in the scaling analysis.

\section{Conclusions and Outlook}
\label{sec:Conclusions}

In this paper, we have applied the UER framework to the determination of the QCD crossover line at finite baryon chemical potential $\hmuB$. Using $(2+1)$-flavor lattice QCD simulations with physical quark masses on a $32^3\times8$ lattice, we reconstruct the baryon number susceptibility $\dimensionlesssusctwo$ and determine the pseudocritical temperature $\pseudocritical$ from the maxima of its thermal derivatives $\dimensionlessch_T$ and $\dimensionlessch_C$ as functions of $T$.
We determine the leading curvature coefficient $\curtwo$ using two distinct approaches: i) a direct analysis of the pseudocritical line and ii) scaling relations evaluated at $\hmuB=0$. From the $\hmuB$ dependence of the pseudocritical temperatures defined from the peaks of $\dimensionlessch_T$ and $\dimensionlessch_C$, we obtain
\begin{equation*}
\curjusttwocrt^{B,T}=0.029(32) \quad,\quad
\curjusttwocrt^{B,C}=0.033(27)\;.
\end{equation*}
These direct determinations are substantially less constrained than the corresponding estimates obtained from the scaling relations, but remain consistent with them within the statistical uncertainties. The scaling-based analysis gives
\begin{align*}
\curjusttwoscl^B(\pseudocritical^T)&=0.02082(14),\\
\curjusttwoscl^B(\pseudocritical^C)&=0.02064(15),\\
\curjusttwoscl^B(\pseudocritical^\chi)&=0.02075(14),
\end{align*}
where $\pseudocritical^\chi$ is determined from the peak position of the fourth-order baryon number susceptibility $\dimensionlesssuscfour$. The scaling-based curvature estimates are obtained from ratios involving $\dimensionlesssuscfour$ and the thermal derivatives of $\dimensionlesssusctwo$, for which the non-universal scaling parameters cancel in the scaling regime. Even though the former outlines estimates of $\kappa_2^B$, $\curjusttwocrt^T$ and $\curjusttwocrt^C$ that are less constrained along with relatively larger uncertainties, the consistency of these estimates with the same derived from scaling relations, $\curjusttwoscl^{B,T}$ and $\curjusttwoscl^{B,C}$ provides a nontrivial cross-check of the resummed description, as well as of these two complementary approaches of determining $\curtwo$.

Within the present analysis, we have explicitly investigated the dependence of the pseudocritical temperatures, $\pseudocritical^T$ and $\pseudocritical^C$ on the order of truncation in UER, interpolating functional form, temperature interpolation in form of numerical resolution of the cubic spline fit procedure as well as also on the range of temperature utilised for fitting the data points. As vividly illustrated in Appendix \ref{appsec:systematic-err-an}, these arguments are presented to validate our choice of the parameters of fitting considered for this work and the analysis. 
Statistical correlations among derived observables are retained through the bootstrap procedure, with quantities entering derived estimators are evaluated on the same bootstrap samples before forming ratios, derivatives, or pseudocritical quantities. Thus, the quoted statistical uncertainties do account for the correlations among the underlying observables rather than treating them as statistically independent.
In particular, the UER approach independently reconstructs the finite-density susceptibility and its associated thermal derivatives, while the resulting observables satisfy the scaling relations for the curvature underlying corresponding Taylor-expansion analyses. This demonstrates consistency of the resummation framework with previous lattice QCD determinations based on Taylor expansions. The present results demonstrate that UER can be extended beyond bulk thermodynamic observables to probe the location and curvature of the QCD crossover line at finite density. Together with our previous studies of the resummed equation of state \cite{Mitra:2023unb,Mitra:2023phd,Mitra:2022cum,Mitra:2022Lat,Mitra:2022dae} and Lee-Yang zeros \cite{Mitra:2024czm}, this work establishes the framework as a complementary approach for investigating finite-density QCD in the vicinity of $\hmuB=0$. 

The present analysis is performed on $32^3\times8$ lattices, corresponding to an aspect ratio $N_s/N_\tau=4$, and the value $m_\pi L\simeq3.3$. Previously related HISQ studies on lattices with similar and comparable aspect ratios have found small finite-volume effects on bulk thermodynamic observables and pseudocritical temperatures within their quoted uncertainties. These results do provide useful guidance, but nevertheless a dedicated finite-volume study of these observables using UER would provide an important consistency check and a direct assessment of the associated systematic effects.
A direct determination of finite-volume effects would require calculations at additional spatial volumes at fixed lattice spacing, over the full temperature range considered here. Such volumes with comparable statistics are not yet available for the present analysis. We therefore leave this systematic investigation of finite-volume effects using multiple spatial volumes for a future work.
Regarding cutoff effects, the present study employs lattice with a single temporal extent, $N_\tau=8$. Consequently, no continuum extrapolation is performed and residual discretization effects are not removed. A quantitative determination of these effects requires calculations at multiple values of $N_\tau$. Previous HISQ studies indicate that cutoff effects in thermodynamic observables are reduced toward larger $N_\tau$, while comparisons with continuum-extrapolated results suggest that discretization effects at $N_\tau=8$ can lead to shifts of a few MeV in pseudocritical temperatures. Such comparisons provide an indication of the possible size of the effect but do not constitute a continuum extrapolation of the present calculation. The results presented here should therefore be regarded as $N_\tau=8$ results. Dedicated calculations at finer lattice spacings are required for a controlled continuum determination.

A more precise determination of the curvature from the pseudocritical-line approach requires improved control over the definition of the pseudocritical temperature at finite $\hmuB$, together with higher statistics and a systematic assessment of alternative pseudocritical estimators. The inclusion of higher-order terms in the parametrization of the crossover line also becomes important once the statistical precision is sufficient to constrain the lower-order terms. While the peak positions of $\dimensionlessch_T$ and $\dimensionlessch_C$ provide well-defined pseudocritical estimators at $\hmuB=0$, their interpretation at finite $\hmuB$ is not uniquely established, the reasons of which have also been highlighted in this work. Further theoretical and numerical studies of their finite-density behavior are therefore needed.
Future studies with higher-order cumulants, finer lattice spacings and multiple spatial volumes for analysing cut-off and finite-volume effects will enable controlled continuum and infinite-volume limit studies and provide a more quantitative assessment of the systematic uncertainties. Exploring the quark-mass dependence of these observables within the UER framework and approaching chiral limit also remains an important direction for future work.
All relevant data for these computations depicted in the figures are in Ref.\cite{mitra_2026_23356418}.

\vspace*{.6cm}

\section*{Acknowledgements}

I sincerely acknowledge useful discussions with Frithjof Karsch and Jishnu Goswami in this regard. 
This work has been supported by the Deutsche Forschungsgemeinschaft
(DFG, German Research Foundation) Project No. 315477589-TRR 211. 
Computations and data analysis required for this work have been performed on the GPU cluster at Bielefeld University, Germany

\bibliography{bibliography}

\appendix
\onecolumngrid

\section{Singular dependence of \texorpdfstring{$\dimensionlessch_2^B$}{chi2B} and its \texorpdfstring{$\hmuB$}{hmuB}-dependent correction}
\label{appsec:pseudocritical-scaling-analysis}

Starting from the singular part of the free-energy density,
$$
\hat{f}_s^{\rm sing}=-\hat{p}_s^{\rm sing}
=h_0\,h^{1+1/\delta}f_f(z),
$$
where \(z=t\,h^{-1/\beta\delta}\), the reduced temperature and symmetry-breaking field are given by
$$
t=\frac{1}{t_0}\left(\frac{T-T_c}{T_c}
+\kappa_2\hat\mu_B^2+\mathcal{O}(\hat\mu_B^4)\right),
\qquad
h=\frac{H}{h_0}.
$$
Here, $t_0$ and $h_0$ are non-universal constants. The singular contribution to the second-order baryon susceptibility, $\dimensionlessch_2^B = \partial^2 \hat{p} / \partial \hmuB^2$ can then be written as
\begin{equation}
\dimensionlessch_2^{B,s}
=
h_0\left[
f_f''(z)\frac{4\kappa_2^2}{t_0^2}
h^{-\alpha/(\beta\delta)}\hat\mu_B^2
+
f_f'(z)\frac{2\kappa_2}{t_0}
h^{(1-\alpha)/(\beta\delta)}
\right],
\end{equation}
where the hyperscaling relation corresponding to three dimensional universality class, \(2-\alpha=\beta(1+\delta)\) has been used. Factoring out the \(\hat\mu_B\)-independent prefactor, this takes the form
\begin{equation}
\dimensionlessch_2^{B,s}
=
A\left[
f_f'(z)
+
\frac{2\kappa_2}{t_0}
h_0^{1/(\beta\delta)}
H^{-1/(\beta\delta)}
\hat\mu_B^2 f_f''(z)
\right],
\end{equation}
where $ A=2\kappa_2h_0 h^{(1-\alpha)/\beta\delta} / t_0$ and $z_0=h_0^{1/\beta\delta}/t_0$. Thus on simplifying the expressions, we find
\begin{equation}
\chi_2^{B,s}
=
A\left[
f_f'(z)
+
2\kappa_2z_0H^{-1/(\beta\delta)}
\hat\mu_B^2 f_f''(z)
\right].
\end{equation}

On using the scaling estimate of \(\kappa_2=0.02\), \(H=1/27\) and \(z_0=1.42\) \cite{Ding:2024sux}, where the critical exponents of the three-dimensional \(O(2)\) universality class $\beta=0.34864, \delta=4.7798$ \cite{Karsch:2023pga} are used, we find the coefficient of the correction term, \(\hat\mu_B^2 f_f''(z)\) leading in $\Ob(\hmuB^2)$ is
$$
2\kappa_2z_0H^{-1/(\beta\delta)}
\simeq 0.41,
$$
and hence
\begin{equation}
\dimensionlessch_2^{B,s} = 
A\left[
f_f'(z)+0.41\,\hat\mu_B^2 f_f''(z)
\right].
\end{equation}
Taking a temperature derivative $\dimensionlessch_T^s=T\,\partial \dimensionlessch_2^s / \partial T$ and $\dimensionlessch_C^s = T_c\,\partial \dimensionlessch_2^s / \partial T$, the corresponding singular contributions to these thermal-derivative observables have the schematic form
\begin{equation}
\dimensionlessch_T^{s} = A_T\,\left[f_f''(z)+0.41\,\hat\mu_B^2f_f'''(z)\right]
\quad,\quad
\dimensionlessch_C^{s} = A_C\,\left[f_f''(z)+0.41\,\hat\mu_B^2f_f'''(z)\right] \;,
\end{equation}
where the overall positive proportionality factors, $A_T$ and $A_C$ differ for \(\chi_T\) and \(\chi_C\). 

A quantitative determination of this correction would require a reliable evaluation of the ratio \(f_f'''(z)/f_f''(z)\) using the appropriate \(O(2)\) scaling function in the range of \(z\) relevant for the present analysis.
To estimate this range at \(\hat\mu_B=0\), we make use of \(T_c=143.7\) MeV, \(H=1/27\), and \(z_0=1.42\). With \(T_{\rm pc}^T=160.2\) MeV and \(T_{\rm pc}^C=158.1\) MeV, one obtains
\begin{equation}
z_{\rm pc}^T
= z\,(\pseudocritical^T) =
z_0\,\frac{T_{\rm pc}^T-T_c}{T_c}\,
H^{-1/(\beta\delta)}
\simeq1.17 \quad,\quad 
z_{\rm pc}^C
= z\,(\pseudocritical^C) =
z_0\,\frac{T_{\rm pc}^C-T_c}{T_c}\,
H^{-1/(\beta\delta)}
\simeq1.02\;.
\label{eq:z-at-muB=0}
\end{equation}
At \(\hat\mu_B=0.5\), using the same value \(\kappa_2=0.02\), these become approximately
\begin{equation}
z_{\rm pc}^T
=
z_0\,\left(\frac{T_{\rm pc}^T-T_c}{T_c} +  \kappa_2\,\hmuB^2\right)\,
H^{-1/(\beta\delta)}
\simeq1.22 \quad , \quad
z_{\rm pc}^C
=
z_0\,\left(\frac{T_{\rm pc}^C-T_c}{T_c} + \kappa_2\,\hmuB^2\right)\,
H^{-1/(\beta\delta)}
\simeq1.07\;.
\label{eq:z-at-muB not 0}
\end{equation}
Thus close to the crossover temperature and with our working range of $\hmuB$, $\hmuB \in [0:0.5]$ considered here with $O(2)$ universal scaling relations, the relevant values of \(z\) remain in the range, \(z\simeq1\)--\(1.2\). 
We do not have such a controlled \(O(2)\) determination here, due to which we refrain from assigning a numerical value to this ratio. Nevertheless, the explicit \(\hat\mu_B^2\) pre-factor provides a useful indication of the expected size of the correction. 
In particular, at the largest value considered in this work, \(\hat\mu_B=0.5\), this has a magnitude $\sim 
0.41\,\hat\mu_B^2=0.1025 \sim 0.1$.
Thus unless the ratio \(f_f'''/f_f''\) becomes quite large in the relevant scaling region of $z$ values given above in Eqns.\ref{eq:z-at-muB=0} and \ref{eq:z-at-muB not 0}, the higher-derivative contribution remains parametrically suppressed.

The available \(O(4)\) scaling-function analysis in Ref.~\cite{Schmidt:2015kea} indicates that \(f_f'''/f_f''\) remains of moderate magnitude $\sim 0.3 - 0.5$ in this region, with \(f_f''\) and \(f_f'''\) having the same sign. While this observation cannot be directly interpreted as a quantitative estimate for the \(O(2)\) theory, it provides a  qualitative guidance that the quark-mass dependent higher-derivative term need not become large for $O(2)$ in particularly in the corresponding crossover region in case of physical quark masses, $H=1/27$. This effect gets explicitly suppressed by the leading coefficient \(0.41\hat\mu_B^2 \sim 0.1 \) at $\hmuB=\hmuB^{max}=0.5$ and it is evident, that this suppression is more for lower $\hmuB < 0.5$. Together with the smooth behavior expected for the universal scaling functions in the relevant \(z\) interval, this provides an indication that the finite-\(\hat\mu_B\) modification of the thermal-derivative peak condition can be sub-leading for \(\hat\mu_B\leq0.5\).

This provides a justification for using the maxima of \(\dimensionlessch_T\) and \(\dimensionlessch_C\) as practical proxy estimators for the pseudocritical temperature, denoted by \(T_{\rm pc}^T\) and \(T_{\rm pc}^C\), respectively, in the small-\(\hat\mu_B\) region considered here. We stress, however, that this argument is not a quantitative proof that the correction is negligible, since the required \(O(2)\) ratio \(f_f'''/f_f''\) has not been independently determined in the present work.
At \(\hat\mu_B=0\), the derivative-based definitions provide natural estimators of the crossover location. At nonzero \(\hat\mu_B\), however, additional derivatives of the scaling function enter the peak condition, as illustrated above. Consequently, \(T_{\rm pc}^T\) and \(T_{\rm pc}^C\) should be regarded here as practical proxy or approximate definitions of the pseudocritical temperature rather than exact determinations of the finite-density pseudocritical line.
Within the range \(0\leq\hat\mu_B\leq0.5\), the explicit \(\hat\mu_B^2\) suppression of the additional scaling contribution and the absence of any indication of a rapidly varying higher-derivative contribution in the relevant crossover region support this approximate interpretation. A more precise assessment of the systematic difference between these proxy definitions and the exact finite-density pseudocritical condition would require a dedicated analysis of the \(O(2)\) free-energy scaling function and its higher derivatives, which is beyond the scope of present work.

\section{Dependence of pseudocritical temperature estimate and extraction procedure}
\label{appsec:systematic-err-an}

In this appendix section, we discuss the dependence of underlying systematics related to the determination of $\pseudocritical^T$ and $\pseudocritical^C$ on several factors : resummation order, functional form used for fitting data points, temperature interpolation in form of numerical resolution of intermediate points between a given range of temperature, the fitting range of temperature used case-wise for cubic splines, cubic and quartic global polynomials.

\subsection{Resummation order}

As a first assessment of systematic uncertainties, we investigate the dependence of the pseudocritical temperatures on the truncation order of the UER. As shown in Table~\ref{Tab:Resummation-order}, the pseudocritical temperatures obtained from the second- and fourth-order resummed susceptibilities are in good agreement over the range \(0\leq\hat\mu_B\leq0.5\). For \(\hat\mu_B\lesssim0.3\), the differences between the two truncation orders are at most a few tenths of a MeV for both pseudocritical estimators. At the largest values of \(\hat\mu_B\), the central values show a somewhat larger difference, reaching about \(1.1\) MeV for \(\pseudocritical^T\) and \(0.7\) MeV for \(\pseudocritical^C\) at \(\hat\mu_B=0.5\). These differences remain comparable to or smaller than the statistical uncertainties of the corresponding fourth-order determinations, and are therefore statistically consistent within the present precision.

\begin{table}[htbp]
\begin{tabular}{|c| c c || c | c c || c  c |}
     \hline
     \multicolumn{3}{|c||}{2nd order} & \multicolumn{3}{|c||}{4th order} & \multicolumn{2}{|c|}{Difference}\\
     \hline\hline
     $\hmuB$ & $\pseudocritical^T$ & $\pseudocritical^C$ & $\hmuB$ & $\pseudocritical^T$ & $\pseudocritical^C$ 
     & $\Delta_{\rm t} \pseudocritical^T$ & $\Delta_{\rm t} \pseudocritical^C$ \\
     \hline\hline
      0.0 & 160.2(4)(3) & 158.1(5)(3)  &   0.0 & 160.2(4)(3)  & 158.1(5)(3) & 0.0 & 0.0 \\
      0.1 & 160.0(4)(3) & 157.9(5)(4) &   0.1 & 160.1(5)(3)  & 158.1(5)(3) & 0.1 & 0.2 \\
      0.2 & 159.8(5)(4) & 157.7(5)(4) &  0.2 & 160.0(5)(2)  & 158.0(5)(1) & 0.2 & 0.3 \\
      0.3 & 159.5(5)(3) & 157.5(5)(3) &  0.3 & 159.8(8)(1)  & 157.7(6)(0) & 0.3 & 0.2 \\
      0.4 & 158.5(8)(7) & 156.5(5)(5) &  0.4 & 159.2(14)(0)  & 157.0(13)(0) & 0.3 & 0.5 \\
      0.5 & 157.8(12)(8) & 155.9(9)(5) &  0.5 & 158.9(29)(1) & 156.6(26)(0) & 1.1 & 0.7 \\
      \hline 

\end{tabular}
\caption{Dependence of the pseudocritical temperatures $\pseudocritical^T$ and $\pseudocritical^C$ on the truncation order of the UER. Results obtained using the second- and fourth-order resummations are shown for several values of $\hmuB$.}
 \label{Tab:Resummation-order}
\end{table}

The observed truncation-order dependence is consequently small compared with the overall uncertainties, although it becomes more visible at larger \(\hat\mu_B\), where the statistical uncertainties also increase. We therefore use the fourth-order UER results for the remainder of the systematic analysis. The comparison with the second-order resummation provides an estimate of the sensitivity of the extracted pseudocritical temperatures to the truncation order.

\subsection{Fit form or function dependence}

\begin{table}[htbp]
    \centering
    \begin{tabular}{|c || c | c | }
    \hline
    Type of fit & $\pseudocritical^T$ & $\pseudocritical^C$ \\
    \hline
       Cubic polynomial & 160.4(2) & 158.3(1)  \\
       \hline
       Quartic polynomial & 160.7(2) & 158.7(3) \\
       \hline
       Cubic spline & 160.2(4) & 158.1(6) \\
       \hline
    \end{tabular}
    \caption{Dependence of the pseudocritical temperatures on the fitting or interpolation form for $135 \leq T \leq 172$ MeV. The cubic spline is used for the central analysis. The spread among the different fit forms involving cubic and quartic polynomials as global fits, is used to assess the associated systematic uncertainty.}
    \label{tab:fit-form}
\end{table}
We further investigate the dependence of the pseudocritical temperatures on the functional form used to represent the temperature dependence of the observables. Table~\ref{tab:fit-form} compares the results obtained using a cubic polynomial, a quartic polynomial, and a cubic spline over the range $135\leq T\leq172$ MeV. The three choices give mutually consistent values within the combined uncertainties. Relative to the cubic-spline result, the largest shifts are $0.5$ MeV for $T_{\rm pc}^T$ and $0.6$ MeV for $T_{\rm pc}^C$, as observed from this Table \ref{tab:fit-form}. These variations provide an estimate of the sensitivity of the pseudocritical temperature to the choice of interpolation, or the fitting form, given other parameters remain the same.

\subsection{Temperature interpolation}

\begin{table}[htbp]
    \centering
    \begin{tabular}{|c || c | c | }
    \hline
    ${\rm N}_{\rm Fine}$ & $\pseudocritical^T$ & $\pseudocritical^C$ \\
    \hline
       2 & 172.000(0)(0) & 172.000 (0) (0)  \\
       \hline
       20 & 160.153 (416) (290) & 158.134 (480) (309) \\
       \hline
       200 & 160.163 (428) (291) & 158.108 (533) (346) \\
       \hline
       1000 & 160.163 (428) (291) & 158.108 (533) (346) \\
       \hline       
       2000 & 160.163 (428) (291) & 158.108 (533) (346) \\
       \hline
    \end{tabular}
    \caption{Convergence of the pseudocritical temperatures with the number of fine points, ${\rm N}_{\rm Fine}$, used in the spline interpolation. The results are stable for ${\rm N}_{\rm Fine}\geq 200$, demonstrating that the numerical resolution of the interpolation does not contribute appreciably to the quoted uncertainties.}
    \label{tab:temperature-interpolation}
\end{table}

To verify that the determination of the pseudocritical temperatures is insensitive to the numerical resolution of the spline interpolation, we vary the number of fine points, ${\rm N}_{\rm Fine}$, used to locate the extremum of the interpolated functions for $\hmuB=0$. 
In this analysis outlined here, we restrict the range of temperature identical to the central analysis of this work, $[135:172]$ MeV, and use cubic spline interpolating fits for fitting the data points in this range, with $2000$ bootstrap samples. As mentioned before, we outline this analysis here only for fourth order UER.
The relevant results are summarized in Table~\ref{tab:temperature-interpolation}. Both $T_{\rm pc}^T$ and $T_{\rm pc}^C$ show clear numerical convergence with increasing ${\rm N}_{\rm Fine}$. In particular, increasing ${\rm N}_{\rm Fine}$ from 20 to 200 changes the central values by only $0.01$ MeV for $T_{\rm pc}^T$ and $0.025$ MeV for $T_{\rm pc}^C$, while the results for ${\rm N}_{\rm Fine}=200,1000$ and $2000$ are identical. Thus, the numerical resolution of the interpolation is negligible compared with the statistical and interpolation-range uncertainties. We therefore use ${\rm N}_{\rm Fine}=2000$ in the subsequent analysis, involving cubic spline interpolating fitting to the respective data points.

The cubic spline is adopted as our central interpolation because it provides a smooth representation of the temperature dependence without imposing a specific global polynomial form. The observed fit-form dependence is included in our analysis of the systematic uncertainty associated with the temperature interpolation.

\subsection{Fitting range of temperature}

\begin{table}[H]
\centering
\begin{tabular}{|c | c||c c|c|}
     \hline
     \multicolumn{5}{|c|}{Cubic Spline} \\
     \hline\hline
     $T_{\min}$ & $T_{\max}$ & $\pseudocritical^T$ & $\pseudocritical^C$ &
     \# Data points \\
     \hline\hline
      130.0 & 172.0 & 160.4(7)(4) & 158.4(8)(7) & 8 \\     
      135.0 & 172.0 & 160.2(4)(3) & 158.1(5)(3) & 7 \\
      145.0 & 172.0 & 160.2(3)(2) & 158.2(4)(2) & 6 \\
      150.0 & 172.0 & 160.0(3)(0) & 157.6(5)(0) & 5 \\
     \hline\hline
      135.0 & 180.0 & 157.4(80)(41) & 155.3(78)(41) & 8 \\
      135.0 & 172.0 & 160.2(4)(3) & 158.1(5)(3) & 7 \\
      135.0 & 170.0 & 160.1(4)(3) & 158.1(6)(4) & 6  \\
      135.0 & 165.0 & 160.0(12)(11) & 158.2 (8)(7) & 5 \\      
      \hline 
\end{tabular}
\caption{Table showing dependence of $\pseudocritical$ obtained from the peak locations of $\dimensionlessch_x, x\equiv T,C$, on the temperature interval $[T_{\min},T_{\max}]$ used for the interpolation. The first group probes broader interpolation intervals, the second varies the lower boundary at fixed $T_{\max}=172$ MeV, and the third varies the upper boundary at fixed $T_{\min}=135$ MeV.} 
\label{tab:Tpc-Fitrangefromboth-cubic splines}
\end{table}

We investigate the dependence of the pseudocritical temperatures on the temperature interval used for the cubic-spline interpolation, and the relevant results are outlined in Table \ref{tab:Tpc-Fitrangefromboth-cubic splines}. First, we vary the lower bound \(T_{\min}\) while keeping \(T_{\max}=172\) MeV fixed. The resulting values of \(\pseudocritical^T\) and \(\pseudocritical^C\) show no significant systematic variation as the lower-temperature data points are removed. Similarly, fixing \(T_{\min}=135\) MeV and varying the upper bound between \(165\) and \(172\) MeV results in essentially unchanged central values, with the statistical uncertainties increasing as the number of interpolation points is reduced. These results indicate that the determination of the pseudocritical temperatures is stable against reasonable variations of the interpolation range.

For comparison, extending the interpolation interval to \(T_{\max}=180\) MeV leads to substantially larger uncertainties and a noticeable shift in the extracted central values. This behavior reflects the increased sensitivity of the spline interpolation to the high-temperature boundary, where the available data provide weaker constraints. We therefore regard the \(135\)--\(180\) MeV result as an indication of boundary sensitivity rather than as evidence for a systematic shift of the pseudocritical temperature.

\begin{table}[H]
\centering
\begin{tabular}{|c | c|| c c|||c | c|| c  c|}
     \hline
     \multicolumn{4}{|c|||}{Cubic polynomial} & \multicolumn{4}{|c|}{Quartic polynomial} \\
     \hline
     $T_{min}$ & $T_{max}$ & $\pseudocritical^T$ & $\pseudocritical^C$ & $T_{min}$ & $T_{max}$ & $\pseudocritical^T$ & $\pseudocritical^C$ \\
     \hline\hline
      130.0 & 172.0 & 161.4(2) & 158.8(1) &                              130.0 & 172.0 & 161.0(1) & 159.3(1) \\    
      135.0 & 172.0 & 160.4(2) & 158.3(1) &
      135.0 & 172.0 & 160.7(2) & 158.7(3) \\
      145.0 & 172.0 & 160.4(1) & 158.4(1) &                              145.0 & 172.0 & 159.6(4) & 157.6(4) \\
      150.0 & 172.0 & 160.0(3) & 157.6(5) &                              150.0 & 172.0 & 160.2(8) & nan \\
     \hline\hline
      135.0 & 180.0 & 160.70(6) & 158.50(5) &                            135.0 & 180.0 & 160.6(2) & 158.4(2) \\
      135.0 & 172.0 & 160.4(2) & 158.3(1) & 
      135.0 & 172.0 & 160.7(2) & 158.7(3) \\
      135.0 & 170.0 & 160.6(4) & 158.4(3) & 
      135.0 & 170.0 & 160.2(3) & 158.7(2)  \\
      135.0 & 165.0 & 161.8 (10) & 159.4(8) & 
      135.0 & 165.0 & 158.3(5)  & 157.4(3) \\      
      \hline 
\end{tabular}
\caption{Dependence of the pseudocritical temperatures, obtained from the peak locations of $\dimensionlessch_x$, $x\equiv T,C$, on the temperature interval $[T_{\min},T_{\max}]$ used in the cubic and quartic polynomial fits. The first group varies the lower boundary at fixed $T_{\max}=172$ MeV, while the second varies the upper boundary at fixed $T_{\min}=135$ MeV.}
\label{tab:Tpccubic-Fitrangefromboth-cubic and quartic polynomials}
\end{table}

We also examine the dependence of the pseudocritical temperatures on the fit range using global cubic and quartic polynomial forms. As depicted by Table \ref{tab:Tpccubic-Fitrangefromboth-cubic and quartic polynomials} for the cubic polynomial, the results remain relatively stable when the lower boundary is varied from \(135\) to \(145\) MeV at fixed \(T_{\max}=172\) MeV, while a larger variation is observed when the fit is extended to \(T_{\min}=130\) MeV or restricted to \(T_{\min}=150\) MeV. At fixed \(T_{\min}=135\) MeV, the \(135\)--\(170\) and \(135\)--\(172\) MeV fits give consistent results, whereas the narrower \(135\)--\(165\) MeV interval leads to substantially larger uncertainties.
The quartic polynomial exhibits a stronger dependence on the fit range. In particular, reducing the number of data points leads to increased variation of the extracted pseudocritical temperatures and larger uncertainties, and the quartic fit does not provide a determination of \(\pseudocritical^C\) for the \(150\)--\(172\) MeV interval. The \(135\)--\(165\) MeV quartic fit also shows a substantial shift relative to the broader fit ranges. These observations indicate that higher-order global polynomial forms become increasingly sensitive to the available fit interval when the number of data points is small.

\end{document}